\documentstyle[prd,aps,preprint,tighten,epsfig]{revtex}

\begin{document}

\draft
\preprint{\begin{tabular}{r}
{\bf hep-ph/0008018} \\
{~}
\end{tabular}}

\title{Possible Implications of Small or Large CP Violation \\
in $B^0_d$ vs $\bar{B}^0_d \rightarrow J/\psi K_{\rm S}$ Decays}
\author{\bf Zhi-zhong Xing}
\address{Institute of High Energy Physics, 
P.O. Box 918, Beijing 100039, China 
\footnote{Mailing address} \\
(Electronic address: xingzz@mail.ihep.ac.cn)}
\maketitle

\begin{abstract}
We argue that a small or large CP-violating asymmetry ${\cal A}_{\psi K_S}$
in $B^0_d$ vs $\bar{B}^0_d\rightarrow J/\psi K_S$
decays, which seems to be favored by the recent BaBar or Belle data,
might hint at the existence of new physics in $B^0_d$-$\bar{B}^0_d$ mixing. 
We present a model-independent framework to show how new physics in 
$B^0_d$-$\bar{B}^0_d$ mixing modifies the standard-model CP-violating 
asymmetry ${\cal A}^{\rm SM}_{\psi K_S}$. We particularly emphasize that 
an experimental confirmation of 
${\cal A}_{\psi K_S} \approx {\cal A}^{\rm SM}_{\psi K_S}$ 
must not imply the absence of new physics in $B^0_d$-$\bar{B}^0_d$ mixing. 
\end{abstract}

\newpage

Recently the BaBar and Belle Collaborations have reported their new
measurements of the CP-violating asymmetry in 
$B^0_d$ vs $\bar{B}^0_d \rightarrow J/\psi K_S$ decays:
\begin{equation}
{\cal A}_{\psi K_S} \; = \; \left \{ \matrix{
0.59 \pm 0.14 ({\rm stat}) \pm 0.05 ({\rm syst}) \; , 
~~~ ({\rm BaBar} ~ \cite{BaBar}) \; , \cr\cr
0.99 \pm  0.14 ({\rm stat}) \pm 0.06 ({\rm syst}) \; , 
~~~ ({\rm Belle} ~ \cite{Belle}) \; . ~~}
\right .
\end{equation}
The central values of these two measurements are apparently
different from that of the previous CDF measurement,
${\cal A}_{\psi K_S} = 0.79 \pm 0.42$ \cite{CDF}; 
and they are also different from the result obtained from global 
analyses of the Cabibbo-Kobayashi-Maskawa (CKM) 
unitarity triangle in the standard model, 
${\cal A}^{\rm SM}_{\psi K_S} = 0.75 \pm 0.06$ \cite{Stocchi}. 
In view of the error bars associated with the BaBar and Belle
measurements, it remains too early to claim any serious 
discrepancy between the experimental result and the standard-model 
prediction. Nevertheless, one cannot rule out the possibility of
${\cal A}_{\psi K_S} < {\cal A}^{\rm SM}_{\psi K_S}$ or
${\cal A}_{\psi K_S} > {\cal A}^{\rm SM}_{\psi K_S}$. A small or 
large CP-violating asymmetry in $B_d \rightarrow J/\psi K_S$ decays 
should be a clean signal of new physics beyond the standard model.

The purpose of this Brief Report is two-fold. First, we 
present a model-independent framework to show how new physics in 
$B^0_d$-$\bar{B}^0_d$ mixing may modify
the standard-model quantity ${\cal A}^{\rm SM}_{\psi K_S}$. 
We find that the possible deviation of ${\cal A}_{\psi K_S}$ from
${\cal A}^{\rm SM}_{\psi K_S}$ can fully be described in terms of three
independent parameters, including the magnitude and phase of the new-physics
contribution to $B^0_d$-$\bar{B}^0_d$ mixing. Second,
we point out that the equality
${\cal A}_{\psi K_S} = {\cal A}^{\rm SM}_{\psi K_S}$ itself
must not mean the absence of new physics in $B^0_d$-$\bar{B}^0_d$ mixing.
Indeed there may exist a specific parameter space for
the new-physics contribution to
$B^0_d$-$\bar{B}^0_d$ mixing, in which the value of
${\cal A}_{\psi K_S}$ coincides with that of 
${\cal A}^{\rm SM}_{\psi K_S}$. Hence
measuring the CP-violating asymmetry ${\cal A}_{\psi K_S}$ alone 
is neither enough to test the standard model nor enough to constrain  
the possible new physics in $B^0_d$-$\bar{B}^0_d$ mixing. 

It is well known that the CP asymmetry ${\cal A}_{\psi K_S}$ arises 
from the interplay of the direct decays of $B^0_d$ and $\bar{B}^0_d$ mesons,
the $B^0_d$-$\bar{B}^0_d$ mixing in the initial state, and the 
$K^0$-$\bar{K}^0$ mixing in the final state \cite{Du}:
\begin{equation}
{\cal A}_{\psi K_S} \; =\; - {\rm Im} \left (\frac{q}{p} \cdot 
\frac{V_{cb}V^*_{cs}}{V^*_{cb}V_{cs}} \cdot \frac{q^*_K}{p^*_K} \right ) \; ,
\end{equation}
where $V_{cb}$ and $V_{cs}$ are the CKM matrix elements,
$p$ and $q$ are the $B^0_d$-$\bar{B}^0_d$ mixing parameters,
$p^{~}_K$ and $q^{~}_K$ are the $K^0$-$\bar{K}^0$ mixing parameters,
and the minus sign on the right-hand side of Eq. (2) 
comes from the CP-odd eigenstate $J/\psi K_S$.
In this expression the tiny penguin contributions to the direct
transition amplitudes, which may slightly modify the ratio
$(V_{cb}V^*_{cs})/(V^*_{cb}V_{cs})$ \cite{Xing95}, have been neglected.
Within the standard model 
$q^{~}_K/p^{~}_K \approx 1$, $q/p \approx V_{td}/V^*_{td}$ 
and $(V_{cb}V^*_{cs})/(V^*_{cb}V_{cs}) \approx 1$
are excellent approximations 
in the Wolfenstein phase convention for the CKM matrix \cite{Wol}. 
Therefore one obtains
\begin{equation}
{\cal A}^{\rm SM}_{\psi K} \; \approx\; - {\rm Im} \left ( 
\frac{V_{td}}{V^*_{td}} \right )
\; \approx \; \sin 2\beta \;\; ,
\end{equation}
where $\beta \equiv \arg [-(V^*_{cb}V_{cd})/(V^*_{tb}V_{td})] 
\approx \arg(-V^*_{td})$ is one of the three inner angles of the 
CKM unitarity triangle \cite{PDG}. A recent global analysis of the
quark flavor mixing data and the CP-violating observables 
in the kaon system yields $\sin 2\beta = 0.75 \pm 0.06$ \cite{Stocchi}.

If the measured value of ${\cal A}_{\psi K_S}$ deviates significantly from the
standard-model prediction in Eq. (3), it is most likely that the
$B^0_d$-$\bar{B}^0_d$ mixing phase $q/p$ consists of
unknown new physics contributions. Of course there may also exist
new physics in $K^0$-$\bar{K}^0$ mixing, contributing a non-trivial
complex phase to ${\cal A}_{\psi K_S}$ through $q^{~}_K/p^{~}_K$.
It is quite unlikely that the tree-level $W$-mediated decays of $B^0_d$ and
$\bar{B}^0_d$ mesons are contaminated by any kind of new physics in a
significant way \cite{Nir99}. 

To be specific, we assume that a possible discrepancy
between ${\cal A}_{\psi K_S}$ and ${\cal A}^{\rm SM}_{\psi K_S}$
mainly results from new physics in $B^0_d$-$\bar{B}^0_d$ mixing. We
therefore write down the ratio $q/p$ in terms of the off-diagonal
elements of the $2\times 2$ $B^0_d$-$\bar{B}^0_d$ mixing Hamiltonian:
\begin{equation}
\frac{q}{p} \; =\; \sqrt{\frac{M^*_{12} - i \Gamma^*_{12}/2}
{M_{12} - i\Gamma_{12}/2}} \; 
\end{equation}
with 
\begin{equation}
M_{12} \; =\; M^{\rm SM}_{12} ~ + ~ M^{\rm NP}_{12} \; 
\end{equation}
and $\Gamma_{12} = \Gamma^{\rm SM}_{12}$. Note that 
$|M_{12}| \gg |\Gamma_{12}|$ is expected to hold 
both within and beyond the standard model, thus we
have $q/p \approx \sqrt{M^*_{12}/M_{12}}$ as a good approximation.
The relative magnitude and the phase difference between the new-physics 
contribution $M^{\rm NP}_{12}$ and the standard-model
contribution $M^{\rm SM}_{12}$ are in general unknown. By definition, we 
may take $|M_{12}| = \Delta M/2$, where 
$\Delta M = (0.487 \pm 0.014) ~{\rm ps}^{-1}$ is the experimentally
measured mass difference between two mass eigenstates of $B_d$ 
mesons \cite{Stocchi00}. Then we parametrize $M^{\rm SM}_{12}$, 
$M^{\rm NP}_{12}$ and $M_{12}$ in the following way:
\begin{equation}
\left \{ \matrix{
M^{\rm SM}_{12} \cr\cr
M^{\rm NP}_{12} \cr\cr
M_{12} \cr} \right \} 
\; =\; \left \{ \matrix{
R_{\rm SM} e^{{\rm i} 2\beta} \cr\cr
R_{\rm NP} e^{{\rm i} 2\theta} \cr\cr
e^{{\rm i} 2\phi} \cr} \right \}
\frac{\Delta M}{2} \; ,
\end{equation}
where $R_{\rm SM}$ and $R_{\rm NP}$ are real and positive parameters, 
$\theta$ represents the new-physics phase, 
and $\phi$ denotes the effective (overall) 
phase of $B^0_d$-$\bar{B}^0_d$ mixing. In this case,
$M^{\rm SM}_{12}$, $M^{\rm NP}_{12}$ and $M_{12}$ (or equivalently,
$R_{\rm SM} e^{{\rm i} 2\beta}$, $R_{\rm NP} e^{{\rm i} 2\theta}$ and 
$e^{{\rm i} 2\phi}$)
form a triangle in the complex plane, as illustrated by Fig. 1.
The dual relation between $R_{\rm SM}$ and $R_{\rm NP}$ 
can be expressed as
$$
R_{\rm NP} \; = \; - R_{\rm SM} \cos 2 (\theta - \beta) 
\pm \sqrt{1 - R^2_{\rm SM} \sin^2 2 (\theta - \beta)} \;\; ,
\eqno{\rm (7a)}
$$
and
$$
R_{\rm SM} \; = \; - R_{\rm NP} \cos 2 (\theta - \beta) 
\pm \sqrt{1 - R^2_{\rm NP} \sin^2 2 (\theta - \beta)} \;\; ,
\eqno{\rm (7b)}
$$
which depends only upon the phase difference $(\theta -\beta)$.
There exist two possible solutions for $R_{\rm NP}$ or $R_{\rm SM}$,
corresponding to $(\pm)$ signs on the right-hand side of
Eq. (7). Numerically, $R_{\rm SM} > 0$ and $R_{\rm NP} \geq 0$ 
must hold for either solution. 

The magnitude of $R_{\rm SM}$ can be
calculated in the box-diagram approximation \cite{Sanda}:
\setcounter{equation}{7}
\begin{equation}
R_{\rm SM} \; = \; \frac{G^2_{\rm F} ~ B_B ~ f^2_B ~ M_B ~ m^2_t}{6\pi^2 ~
\Delta M} ~ \eta^{~}_B ~ F(z) ~ |V_{tb}V_{td}|^2 \; ,
\end{equation}
where $G_{\rm F}$ is the Fermi constant, $B_B$ is the ``bag''
parameter describing the uncertainty in evaluation of the hadronic
matrix element 
$\langle B^0_d |\bar{b} \gamma_\mu (1-\gamma_5)d|\bar{B}^0_d\rangle$,
$M_B$ is the $B_d$--meson mass, $f_B$ is the decay constant,
$m_t$ is the top-quark mass, $\eta_B$ denotes the QCD correction factor,
$V_{tb}$ and $V_{td}$ are the CKM matrix elements,
and $F(z)$ stands for a slowly decreasing monotonic function of
$z \equiv m^2_t/M^2_W$ with $M_W$ being the $W$-boson mass.
At present it is difficult to obtain a reliable value for 
$R_{\rm SM}$, because quite large uncertainties may arise from
the input parameters $B_B$, $f_B$ and $|V_{td}|$.
However, $R_{\rm SM}$ is in general expected to be close to unity, 
no matter what kind of new physics is hidden in $B^0_d$-$\bar{B}^0_d$ 
mixing. Note that $R_{\rm SM} = 1$ must not lead to
$R_{\rm NP} =0$. There is another solution,
$R_{\rm NP} = - 2 \cos 2(\theta - \beta)$ with
$\cos 2 (\theta - \beta) \leq 0$, corresponding to
$R_{\rm SM} =1$. On the contrary, $N_{\rm NP} =0$ must result in
$R_{\rm SM} =1$, as indicated by Eq. (7b).

With the help of Eqs. (5) and (6), we 
recalculate the CP-violating asymmetry ${\cal A}_{\psi K_S}$ and arrive
at the following result:
\begin{equation}
{\cal A}_{\psi K_S} \; =\; \sin (2\phi) \; =\; R_{\rm SM} \sin
(2\beta) ~ + ~ R_{\rm NP} \sin (2\theta)  \; .
\end{equation}
Note that $R_{\rm NP}$, $R_{\rm SM}$, $\beta$, and $\theta$ are dependent
on one another through Eq. (7). Of course, $|{\cal A}_{\psi K}| \leq 1$
holds within the allowed parameter space of $R_{\rm NP}$ and $\theta$.
The ratio of ${\cal A}_{\psi K_S}$ to 
${\cal A}^{\rm SM}_{\psi K_S}$ is given as
\begin{equation}
\xi_{\psi K_S} \; \equiv \; \frac{{\cal A}_{\psi K_S}}
{{\cal A}^{\rm SM}_{\psi K_S}} \; \approx \; 
\frac{\sin (2\phi)}{\sin (2\beta)} \; =\;
R_{\rm SM} + R_{\rm NP} \frac{\sin (2\theta)}{\sin (2\beta)} \; .
\end{equation}
In the literature (e.g., Ref. \cite{Stocchi}), the value of
${\cal A}^{\rm SM}_{\psi K_S} \approx \sin 2\beta$ is obtained from 
a global analysis of the experimental data on 
$|V_{ub}/V_{cb}|$, $B^0_d$-$\bar{B}^0_d$ mixing, 
$B^0_s$-$\bar{B}^0_s$ mixing, and CP violation in $K^0$-$\bar{K}^0$
mixing. The key assumption in such analyses is that there is no
new-physics contribution to the $K^0$-$\bar{K}^0$,
$B^0_d$-$\bar{B}^0_d$, and $B^0_s$-$\bar{B}^0_s$ mixing systems.
If new physics does contribute significantly to the heavy
meson-antimeson mixing instead of the light one, 
one has to discard the direct experimental
data on $B^0_d$-$\bar{B}^0_d$ mixing and $B^0_s$-$\bar{B}^0_s$ mixing
in analyzing the CKM unitarity triangle. In this case, the resultant
constraint on $\sin 2\beta$ becomes somehow looser. 
One may observe, from the figures of the CKM unitarity triangle 
in Refs. \cite{Stocchi,PDG}, that $0.6 \leq \sin 2\beta \leq 0.8$
is a quite generous range constrained by current data on
$|V_{ub}/V_{cb}|$ and CP violation in $K^0$-$\bar{K}^0$ mixing.
Given such a generously allowed region for ${\cal A}^{\rm SM}_{\psi K_S}$,
we conclude that $\xi_{\psi K_S} >0$ is definitely assured. 
Using ${\cal A}^{\rm SM}_{\psi K_S} = 0.75 \pm 0.06$ \cite{Stocchi}
for illustration, we obtain
\begin{equation}
\xi_{\psi K_S} \; = \; \left \{ \matrix{
0.79 \pm 0.26 \; , ~~~ ({\rm BaBar}) \; , \cr\cr
1.32 \pm 0.30 \; . ~~~~ ({\rm Belle}) \; . ~~} \right .
\end{equation}
We see that the BaBar measurement seems to 
indicate $\xi_{\psi K_S} <1$, while the Belle measurement seems to
imply $\xi_{\psi K_S} > 1$. If either possibility could
finally be confirmed with more precise experimental data from 
$B$-meson factories, it would be a very clean signal of new physics \cite{NP}.
If the further data of both BaBar and Belle Collaborations turn to
coincide with each other and lead to $\xi_{\psi K_S} \approx 1$,
however, one cannot draw the conclusion that there is no new physics in
$B^0_d$-$\bar{B}^0_d$ mixing.

Now let us show why ${\cal A}_{\psi K_S} = {\cal A}^{\rm SM}_{\psi K_S}$
must not imply the absence of new physics in $B^0_d$-$\bar{B}^0_d$
mixing. Taking $\xi_{\psi K_S} =1$ and using Eq. (7), we 
obtain the following equation constraining the allowed values of
$\theta$:
\begin{equation}
(1 + R_{\rm SM}) \tan^2 2\theta - 2R_{\rm SM} \tan 2\beta \tan 2\theta
- (1 - R_{\rm SM}) \tan^2 2\beta \; =\; 0 \; .
\end{equation}
Then it is straightforward to find out two solutions for $\tan 2 \theta$:
$$
\tan 2\theta \; =\; \tan 2\beta \; , ~~~~~~~~~~~~
\eqno{\rm (13a)}
$$
or
$$
\tan 2\theta \; =\; \tan 2\beta ~ \frac{R_{\rm SM} -1}{R_{\rm SM} + 1} \; .
\eqno{\rm (13b)}
$$
Note that solution (13a) corresponds to $R_{\rm SM} + R_{\rm NP} =1$.
Solution (13b) implies that $|\tan 2\theta| \ll \tan 2\beta$ may
hold, if $R_{\rm SM}$ is remarkably close to 1. 
Although the afore-obtained region of $\theta$ is quite specific, it does 
exist and give rise to $\xi_{\psi K_S} =1$. Therefore, an experimental
confirmation of $\xi_{\psi K_S} \approx 1$ cannot fully rule out the
possibility of new physics hidden in $B^0_d$-$\bar{B}^0_d$ mixing.

Theoretically, the information on 
$R_{\rm NP}$ and $\theta$ can only be obtained from specific models
of new physics (e.g., the supersymmetric extensions of the standard
model \cite{NP}). An interesting possibility is that
the new-physics contribution conserves CP (i.e., 
$\theta =0$ \cite{Lincoln})
and all observed CP-violating phenomena in weak interactions
are attributed to the non-trivial phase in the CKM matrix.
In this scenario, we obtain
\setcounter{equation}{13}
\begin{equation}
{\cal A}_{\psi K_S} \; =\; \sin (2\phi) \; =\; R_{\rm SM} \sin
(2\beta) \; .
\end{equation}
Obviously ${\cal A}_{\psi K_S}/{\cal A}^{\rm SM}_{\psi K_S}
= R_{\rm SM} \leq 1$ is required, in order to understand the present 
BaBar and Belle measurements. 

It becomes clear that the measurement of ${\cal A}_{\psi K_S}$ itself is not
enough to test the self-consistency of the standard model or to
pin down possible new physics hidden in $B^0_d$-$\bar{B}^0_d$ mixing.
For either purpose one needs to study the CP-violating asymmetries in some
other nonleptonic $B$-meson decays, although most of them are not so
clean as $B^0_d$ vs $\bar{B}^0_d\rightarrow J/\psi K_S$ decays
in establishing the relations between the CP-violating observables and
the fundamental CP-violating parameters \cite{HQ}.

In summary, we have discussed possible implications of a small or large
CP-violating asymmetry in $B^0_d$ vs $\bar{B}^0_d\rightarrow J/\psi K_S$
decays. While such an effect could be attributed to new physics
in $K^0$-$\bar{K}^0$ mixing, it is most likely to result from 
new physics in $B^0_d$-$\bar{B}^0_d$ mixing. Model-independently, we
have formulated the basic features of new-physics effects on CP
violation in $B_d\rightarrow J/\psi K_S$. We have also pointed out
that an experimental confirmation of
${\cal A}_{\psi K_S} \approx {\cal A}^{\rm SM}_{\psi K_S}$ 
must not imply the absence of new physics in $B^0_d$-$\bar{B}^0_d$
mixing. An extensive
study of all hadronic $B$-meson decays and CP asymmetries is 
desirable, in order to test the standard model and probe possible 
new physics at some higher energy scales.

\vspace{0.5cm}

The author would like to thank L. Wolfenstein 
for numerous useful discussions.

\newpage

\begin{figure}[t]
\vspace{-1cm}
\epsfig{file=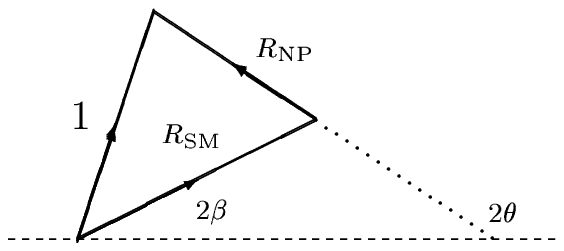,bbllx=6.5cm,bblly=15cm,bburx=15cm,bbury=30cm,%
width=15cm,height=22cm,angle=0,clip=}
\vspace{-13.5cm}
\caption{\small Triangular relation of $M^{\rm SM}_{12}$, $M^{\rm
NP}_{12}$ and $M_{12}$ (rescaled by $\Delta M/2$) in the complex plane.}
\end{figure}

\end{document}